\documentclass{tlp}

\pdfoutput=1

\usepackage[latin1]{inputenc}
\usepackage[T1]{fontenc}
\usepackage{amsfonts}
\usepackage{theorem}
\usepackage{epsfig}
\usepackage{alltt}
\usepackage{macro}

\title[Tracer Driver and Dynamic Analyses of CLP(FD)] {Design and
      Implementation of a Tracer Driver: Easy and Efficient Dynamic
      Analyses of Constraint Logic
      Programs\textsuperscript{\thanks{This work has been partly
      supported by the French RNTL OADymPPaC project and the ERCIM
      fellowship programme.}}}

\author[L. Langevine and M. Ducassé]{Ludovic Langevine\thanks{Most
    of the work has been done while Ludovic Langevine was at INRIA
    Rocquencourt, France}\\
Mission Critical IT, Boulevard de France, 9, Bât A, 1420 -
    Braine-l'Alleud, Belgium \\
\email{llg@missioncriticalit.com}
 \and Mireille Ducassé\\
IRISA/INSA of Rennes, Campus de Beaulieu, 35\,042 Rennes Cedex, France\\
\email{Mireille.Ducasse@irisa.fr}
}

\pagerange{\pageref{firstpage}--\pageref{lastpage}}
\submitted{5 July 2006}
\revised{5 June 2007, 20 December 2007}
\accepted{23 April 2008}

\begin{document}

\long\def\comment#1{}
\maketitle

\label{firstpage}

\begin{abstract}
  Tracers provide users with useful information about program
  executions.  In this article, we propose a ``tracer driver''.  From
  a single tracer, it provides a powerful front-end enabling multiple
  dynamic analysis tools to be easily implemented, while limiting the
  overhead of the trace generation.  
  The relevant execution events are specified by flexible event
  patterns and a large variety of trace data can be given either
  systematically or ``on demand''.  The proposed tracer driver has
  been designed in the context of constraint logic programming;
  experiments have been made within GNU-Prolog.  Execution views
  provided by existing tools have been easily emulated with a
  negligible overhead.  Experimental measures show that the
  flexibility and power of the described architecture lead to good
  performance.

The tracer driver overhead is inversely proportional to the average
time between two traced events. Whereas the principles of the tracer
driver are independent of the traced programming language, it is best
suited for high-level languages, such as constraint logic programming,
where each traced execution event encompasses numerous low-level
execution steps. 
Furthermore, constraint logic programming is especially hard to
debug. The current environments do not provide all the useful dynamic
analysis tools. They can significantly benefit from our tracer driver
which enables dynamic analyses to be integrated  at a very low cost.

To appear in Theory and Practice of Logic Programming (TPLP),
Cambridge University Press.
\end{abstract}
\begin{keywords}
Software Engineering, Debugging, Execution Trace Analysis, Execution
  Monitoring, Execution Tracing, Execution Visualization, Programming
  Environment, Constraint Logic Programming
\end{keywords}

\section{Introduction}

Dynamic program analysis is the process of analyzing program
executions.
It is generally acknowledged that dynamic analysis is complementary to
static analysis; see for example the discussion of
Ball~\cite{ball99}. Dynamic analysis tools include, in particular,
tracers, debuggers, monitors and visualizers.

\begin{figure}[h,h,h]
\begin{center}
  \includegraphics[width=0.75\linewidth]{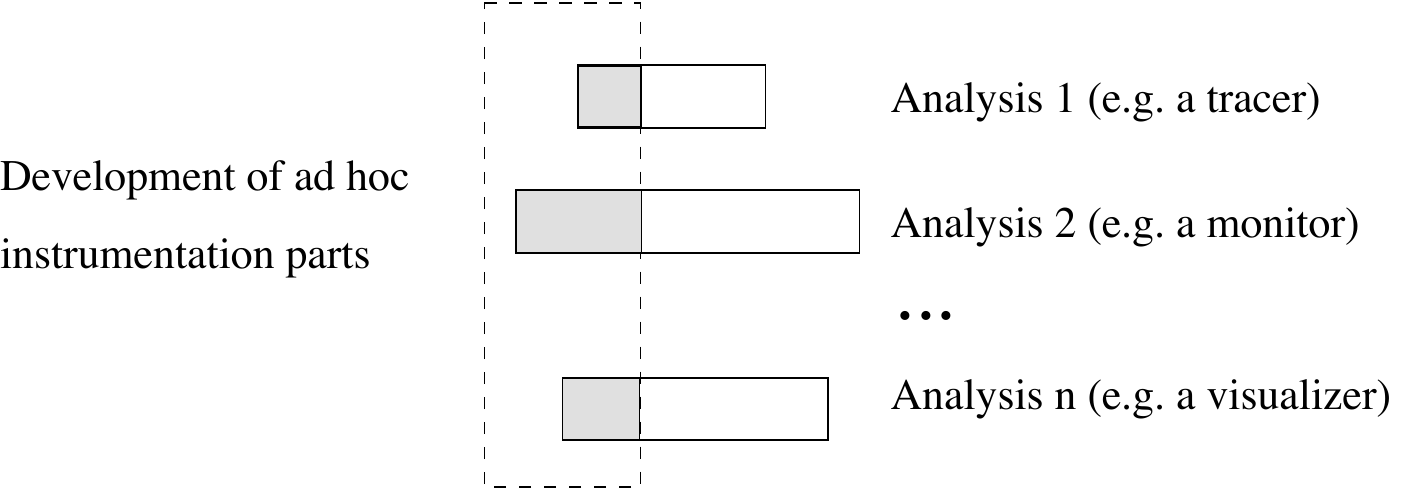}
\end{center}
\vspace{-0.5cm}
\caption{The usual case: all dynamic analysis tools implement a dedicated
  instrumentation part}
\label{fig:duplication}
\end{figure}

In order to be able to analyze executions, some data must be gathered
and some sort of instrumentation mechanisms must be implemented. The
state-of-the-practice, illustrated by Fig.~\ref{fig:duplication}, is
to re-implement the instrumentation for each new dynamic analysis
tool. The advantages are, firstly, that the instrumentation is
naturally and tightly connected to the analysis, and secondly, that it
is specialized for the targeted analysis and produces relevant
information.  The drawback, however, is that this implementation
usually requires much tedious work which has to be repeated by each
tool's writer for each environment. This acts as a brake upon
development of dynamic analysis tools.

\subsection{A Tracer Driver to Efficiently Share Instrumentations} 

In this article we suggest that standard tracers can be used to give
information about executions to several dynamic analysis
tools. Indeed, Harrold et al. have shown that a trace 
consisting of the sequence of program statements traversed as the
program executes subsumes a number of interesting other 
representations such as the set of conditional branches or the set of
paths~\cite{harrold97}.

\begin{figure}[h,h,h]
\begin{center}
  \includegraphics[width=0.75\linewidth]{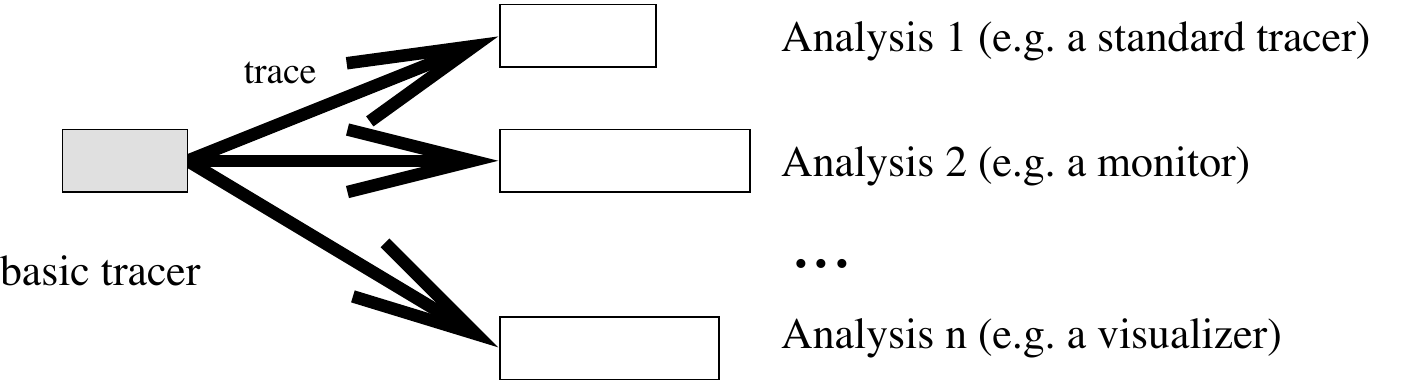}
\end{center}
\vspace{-0.5cm}
\caption{The ``generate-and-dump'' approach: the instrumentation part
  is shared but the amount of data is huge}
\label{fig:generate:and:dump}
\end{figure}

However, the separation of the extraction part from the analysis part
requires care. Indeed, as illustrated by
Fig.~\ref{fig:generate:and:dump}, if execution information is
systematically generated and dumped to the analysis tools, the amount
of information that flows from the tracer to the analysis modules can
be huge, namely several gigabytes for a few seconds of
execution. Whether the information flows through a file, a pipe, or
even main memory, writing such an amount of information takes so much
time that the tools are not usable interactively.  This is especially
critical for debugging, even when it is automated, because users need
to interact in real-time with the tools.

\begin{figure}
\begin{center}
  \includegraphics[width=0.75\linewidth]{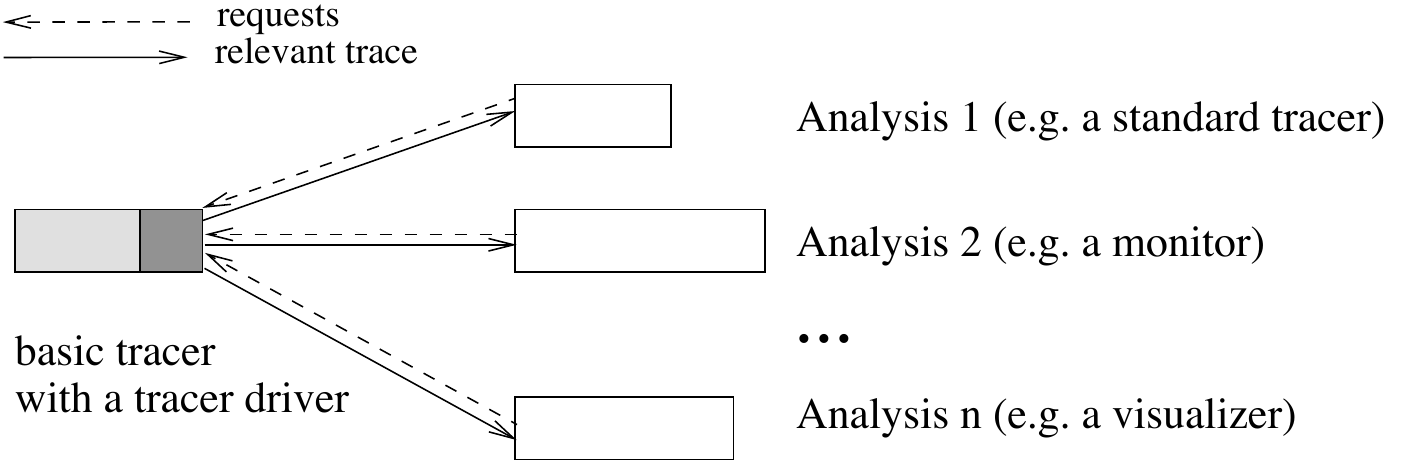}
\end{center}
\vspace{-0.5cm}
\caption{Our supervised generation approach: with a tracer driver only
the relevant part of the execution information is generated}
\label{fig:supervised:generation}
\end{figure}

Reiss and Renieris propose to encode and compact the trace
information~\cite{reiss01}. Their approach is used in a context where
multiple tracing sources send information to the same analysis
module.
In this article, we propose another approach, more accurate when a
single source sends information to (possibly) several analysis
modules. As illustrated by Fig.~\ref{fig:supervised:generation}, we
have designed what we call a ``tracer driver'', whose primary function
is to filter the data on the fly according to requests sent by the
analysis modules. Only the necessary trace information is actually
generated.  This often drastically reduces the amount of trace data,
and significantly improves the performance. 

Therefore, the instrumentation module is shared among several analysis
tools and there is very little slowdown compared to the solution where
each analysis has its dedicated instrumentation.
From a single tracer, the tracer driver provides a powerful
front-end for multiple dynamic analysis tools while limiting the
overhead of the trace generation. 
The consequence is that specifying and implementing dynamic analysis
tools is much easier, without negative impact on the end-user.

\subsection{Interactions between a Tracer and  Analyzers}

In the following, we call  a module that is connected to
a tracer an ``analyzer'' .
In its simplest form the analyzer is only the standard output, or a
file, in which traces are written by a primitive tracer.
Another form of analyzers is traditional debuggers, which are mere
interactive tracers of executions. They handle the interaction once
the execution is stopped at interesting points. They show some trace
information and react on users' commands.
More sophisticated debugging tools exhibit more sophisticated
analyzers.  A trace querying mechanism with a data\-base flavor can be
connected to a tracer and let users investigate executions in a more
thorough way. This has been done for example for C in the Coca
tool~\cite{ducasse99c} and for Prolog in the Opium
tool~\cite{ducasse99}.
A real database can even be used if on the fly performance is not a
big issue. This has been done for a distributed system in
Hy$^+$~\cite{consens94}.
Algorithmic debugging traverses an execution tree in an interactive
way. Users answer queries and an algorithm focuses on nodes which seem
erroneous to users while their children seem correct~\cite{shapiro83}.
Note that the declarative debugger of Mercury is explicitly built on
top of the Mercury tracer~\cite{maclarty05}.
Monitoring tools can be connected to tracers in order to supervise
executions and collect data.  For example, the Morphine tool for
Mercury is able to assess the quality of a test set~\cite{JD02tplp}.
The EMMI tool, for Icon, is able to detect some programming
mistakes~\cite{jeffery94}.
A number of visualization tools use traces to generate graphical
views such as the DiSCiPl views~\cite{deransart2000} for constraint
logic programming.

The latest trend in fault localization consists in mining sets of
program executions to cross check execution traces, see for
example~\cite{jones02,jones05,denmat05b}. 

The interaction modes between the tracer and the analyzers
exhibited by the previous examples are all different and specific.
For primitive tracers, simple visualization and trace mining, the
tracer simply outputs trace information into a given channel. 
Traditional debuggers, trace query systems and declarative debuggers
output information about executions and get user requests.
When information is displayed and until the user sends a request, the
execution is blocked.
Monitors process the trace information on the fly.
They also block the execution until they have finished processing the
current trace information but
without any interaction with users.
At present, all these tools are disjoint and difficult to
merge. Therefore, further mechanisms are required in order to share a
tracer among analyzers of different types.

When the tracer has sent trace information to the analyzers, the above
examples exhibit two behaviors: 1) the execution is blocked waiting for an
answer from the analyzers; this is called ``synchronous interaction''
in the following; 2) the execution proceeds; this is called
``asynchronous interaction'' in the following.
Our tracer driver includes both mechanisms. It enables different
interaction modes between a tracer and analyzers to be integrated in
one single tool. This has several advantages.  Firstly, users do not
switch tools to achieve different aims.  They use a single tool
to trace, debug, monitor and visualize executions.
Secondly, integrating all the possible usages results in a more
powerful tool than the mere juxtaposition of different tools.  For
example, one can, simultaneously, check for known bug patterns, and
collect data for visualization. Whenever a bug is encountered the
tool can switch to a synchronous debugging session, using the
already collected visualization data.  The visualization tool can
also change the granularity of the collected data depending on the
current context.

\subsection{Debugging of Constraint Logic Programs}

The proposed tracer driver has been designed in the context of
constraint logic programming. Experiments have been made within
GNU-Prolog~\cite{gnuprolog}.
Programs with constraints are especially hard to
debug~\cite{meier95}. The numerous constraints and variables involved
make the state of the execution difficult to grasp. Moreover, the
complexity of the filtering algorithms as well as the optimized
propagation strategies lead to a tortuous execution. As a result, when
a program gives incorrect answers, misses expected solutions, or has
disappointing performance, the developer gets very little support
from the current programming environments to improve the program. This
issue is critical because it increases the expertise required to
develop constraint programs.

Some previous papers have addressed this critical issue. Most of them are
based on dynamic analyses. During the execution, some data are
collected in the execution so as to display some graphical views,
compute some statistics and other abstraction of the execution
behavior. Those data are then examined by the programmer to gain a
better understanding of the execution. 
For instance, a display of the search-tree shows users how the search
heuristics behave~\cite{fages02}. Adding some visual clues about the
domain propagation helps users locate situations where propagation is
not strong enough~\cite{carro2000,Outils:OPLstudio}.
A structured inspection of the store helps to investigate constraints
behavior~\cite{goualard2000}.
A more detailed view of the
propagation in specific nodes of the search-tree gives a good insight
to find redundant constraints or select different filtering
algorithms~\cite{simonis2000}. 

A common observation is that there is no ultimate tool that would
meet all the debugging needs. A large variety of complementary tools
already exists, ranging from coarse-grained abstraction of the
whole execution to very detailed views of small sub-parts, and even
application-specific displays.
As a matter of fact, none of the current environments in CLP contain
all of the interesting features  that have already been identified. 
Each of these features requires a dedicated instrumentation of the
execution, or a dedicated annotation of the traced program, to collect
the data they need. Those instrumentations are often hard to make.
Yet, many of those dynamic analyses could be built on top of low-level
tracers. The generated traces can be structured and abstracted by
analyzers in order to produce high-level views.
With our tracer driver, it is easy to develop and explore dynamic
analyses with diverse abstraction levels. For instance, we can first
compute a general view of the search-tree, tracing only the execution
events related to the search-tree construction and ignoring the
propagation events or just computing some basics statistics about
propagation stages. Such an analysis quickly gives a general picture
of the execution. Then, a more specific analysis of, say, a subset of
variables or constraints may provide further details about a sub-part
of the program but may need a more voluminous execution trace.

\subsection{Contributions}

The contributions of this article are threefold.
Firstly, it justifies the need for a tracer driver in order to be able
to efficiently integrate several dynamic analyses within a single
tool. In particular, it emphasizes that both synchronous and
asynchronous communications are required between the tracer and the
analyzer.
Secondly, it describes in breadth and in some depth the mechanisms
needed to implement such a tracer driver: 1) the patterns to specify
what trace information is needed, 2) the language of interaction
between the tracer driver and the analyzers and 3) the mechanisms to
efficiently filter trace information on the fly.
Lastly, an implementation has been achieved inside GNU-Prolog.  The
paper assumes propagation based solvers only for purposes of
exposition. The mechanisms of the tracer driver are independent of the
traced language, they are still applicable to solvers that do not use
propagators.
Experimental measurements on CLP(FD) executions  show that this
architecture increases the trace relevance and drastically speeds up
trace generation and communication. More precisely, the experiments
show that
\begin{enumerate}
\item The overhead of the core tracer mechanisms is small, therefore the
core tracer can be permanently activated
\item The tracer driver overhead is inversely proportional to the average
time between two traced events. It is acceptable for CLP(FD).
\item There is no overhead in the filtering mechanisms when searching
  simultaneously for several patterns.
\item The tracer driver overhead is predictable for given patterns.
\item The tracer driver approach that we propose is more
efficient than sending over a default trace, even to construct
sophisticated graphical views.
\item Answering queries is orders of magnitude more efficient
than displaying traces.
\item There is no need to  restrict the trace information a
  priori. 
\item The performance of our tool is comparable to the state-of-the-practice
while being more powerful and more generic.
\end{enumerate}

Whereas the principles of the tracer driver are independent of the
traced programming language, it is best suited for high-level
languages, such as constraint logic programming, where each traced
execution event encompasses numerous low-level execution steps.

\paragraph{}
In the following, Section~\ref{sec:analyse:tracer_driver} gives an
overview of the tracer driver and in particular the interactions it
enables between a tracer and an analyzer.
Section~\ref{sec:analyse:pattern} specifies the nature of patterns.
Section~\ref{sec:analyse:request} presents the requests that an analyzer
can send to our tracer and how they are processed.
Section~\ref{sec:analyse:filtering} describes in detail our filtering
mechanism and its implementation.
Section~\ref{tracer:implementation:section} discusses the requirements
on the tracer for the overall architecture to be efficient. 
Section~\ref{sec:analyse:performance} gives experimental results and
shows the efficiency of the tracer driver mechanism.
Section~\ref{sec:analyse:related}  discusses related work.

\section{Overview of the tracer driver}
\label{sec:analyse:tracer_driver}
\index{traceur!pilote de}

This section presents an overview of the tracer driver architecture
and, in particular, the interactions it enables between a tracer and 
analyzers. 
The tracer and the analyzers are run as two concurrent processes.
The tracer spies the execution and can output, when needed, certain trace
events with some attached data. Each analyzer listens to the trace and
processes it.

As already mentioned in the introduction, both synchronous and
asynchronous interactions are necessary between the tracer and the
analyzers. On the one hand, some analyzers are highly interactive.
Users may ask for more information about the current event than what
is provided by default. In that cases, it is important that the
execution is blocked until all the analyzers notify it that it can
proceed. On the other hand, if the analyzers passively collect
information there is no need to block the execution.

\begin{figure}[t,t,t]
  \includegraphics[width=\linewidth]{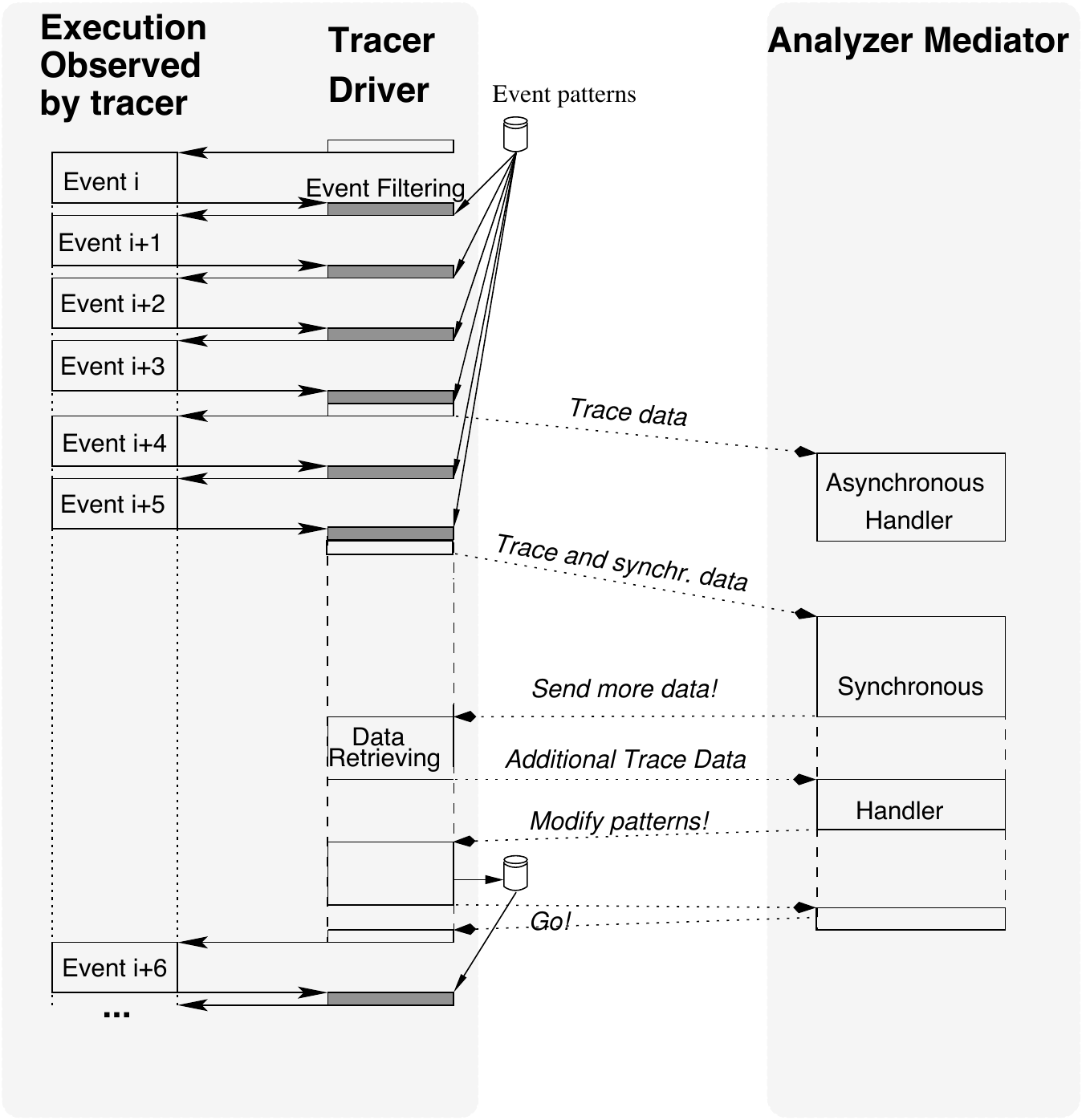}
\vspace{-0.5cm}
\caption{Asynchronous and synchronous interactions between the tracer
  and the analyzer mediator}
\label{fig:coroutine:points}
\end{figure}

An execution trace is a sequence of observed execution events that
have attributes.  The analyzers specify the events to be observed by
the means of \emph{event patterns}. An event pattern is a condition on
the attributes of an event (see details in
Section~\ref{sec:analyse:pattern}).  The tracer driver manages a set
of active event patterns.  Each execution event is checked against the
set of active patterns.  An event matches an event pattern if and only
if the pattern condition is satisfied by the attributes of this event.

An \emph{asynchronous pattern} specifies that, at matching trace events,
some trace data are to be sent to  analyzers without freezing the
execution.
A \emph{synchronous pattern} specifies that, at matching trace events,
some trace data are to be sent to analyzers. The execution is frozen
until the analyzers order the execution to resume. 
An \emph{event handler} is a procedure defined in an analyzer, which
is called when a matching event is encountered.

Fig.~\ref{fig:coroutine:points} illustrates the treatment of the two
types of patterns.  The execution is presented as a sequence of
elementary blocks (the execution events).
An analyzer mediator gathers the patterns requested by the analyzers
and distributes the trace data sent by the tracer driver to the analyzers (see detailed
description Section~\ref{sec:analyse:request}). 
At each trace event, the tracer driver is called to filter the event.
If the current event does not match any of the specified active
patterns, the execution continues (events $i$, $i+1$, $i+2$, $i+4$,
$i+6$). If the current event matches an active pattern, some trace
data are sent to the analyzer mediator (events $i+3$, $i+5$).
If the matched pattern is asynchronous the data is processed by the
relevant analyzer in an asynchronous way (event $i+3$). If the pattern
is synchronous the execution is frozen, waiting for a query from the
analyzer (event $i+5$). The analyzer processes the sent data and can
ask for more data about the state of the execution.  The tracer driver
can retrieve useful data about the execution state and send them to
the analyzer ``on demand''.  The analyzer can also request some
modifications of the active patterns: add new patterns or remove
existing ones.  When no analyzer has any further request to make about
the current event, the analyzer mediator sends the resuming command to
the tracer driver ($go$ command). The tracer then resumes the
execution until the next matching event.

The architecture enables the management of several active patterns.
Each pattern is identified by a label.  A given execution event may
match several patterns. When sending the trace data, the list of
(labels of) matched patterns is added to the trace. Then, the analyzer
mediator calls a specific handler for each matched pattern and
dispatches relevant trace data to it.  If at least one matched pattern
is synchronous, the analyzer mediator waits for every synchronous
handler to finish before sending the resuming command to the tracer
driver.  From the point of view of a given event handler, the
activation of other handlers on the same execution event is
transparent.

This article focuses more on the tracer driver than on the analyzer
mediator. On the one hand, the design and implementation of the tracer
driver is critical with respect to response time. Indeed it is called
at each event and executions of several millions of events (see
Section~\ref{sec:analyse:performance}) are very common. Every
overhead, even the tiniest, is therefore critical. On the other
hand, the implementation of the analyzer mediator is much less
critical because it is called only on matching events. Furthermore,
its implementation is much easier.

\section{Event Patterns}
\label{sec:analyse:pattern}

As already mentioned, an event pattern is a condition on the
attributes of events.  It consists of a logical formula combining
elementary conditions on the attributes.
This section summarizes the information attached to trace events,
specifies the format of the event patterns and gives examples of
patterns.

\subsection{Trace Events}
\label{sec:analyse:trace_event_format}

Some information is attached to each trace event. 
This section summarizes the format of trace event information used in
this article; a more detailed description can be found
in~\cite{langevine04}.
The actual format of trace event information has \emph{no influence}
on the tracer driver mechanisms. The important issue is that events
have attributes and that some attributes are specific to the type of
events.
Note that the pattern language is independent of the traced
language.\\

A constraint program manipulates variables and constraints on these
variables.  Each variable has a \emph{domain}, a finite set of
possible values. The aim of a constraint program is to find a
valuation (or the best valuation, given an objective function) of the
variables such that every constraint is satisfied. To do so,
constraint solvers implement numerous algorithms coming from various
research areas, such as operation research.  Traditionally in logic
programming, the type of events is called a \emph{port}.  There are 14
possible event types in the tracer we use.
Those types of events are partitioned into two classes: control events
and propagation events.
The
control events are related to the handling of the constraint store and
the search:
\begin{itemize}
\item \newVariable{} specifies that a new variable is introduced;
\item \newConstraint{} specifies that the solver declares a new
constraint;
\item \post{} specifies that a declared constraint is introduced into the
store as the active one;
\item \newChild{} specifies that the current solver state corresponds
to a new node of the search-tree;  
\item \jumpTo{} specifies that the solver back-jumps from its current state
to a previous choice-point;
\item \solution{} specifies that the current solver state is a solution (a
toplevel success);
\item \failure{} specifies that the current state is inconsistent.
\end{itemize}
Six ports describe the domain
reductions and the constraint propagation:
\begin{itemize}
\item \reduce{} specifies that a domain is being reduced, this
generates domain updates which have to be propagated;
\item \suspend{} specifies that the active constraint cannot
reduce any more domains and is thus suspended;
\item \entail{} specifies that the active constraint is true;
\item \reject{} specifies that the active constraint is
unsatisfiable;
\item \schedule{} specifies that a domain update is selected
by the propagation loop;
\item \awake{} specifies that a constraint which depends on the
scheduled domain update is awakened;
\item \endoftrace{} notifies the end of the tracing process.
\end{itemize}

\subsection{Event Attributes}

Each event has common and specific attributes. Attributes are
data about the execution event. The common attributes are: the
port, a chronological event number, the depth of the current node in
the search-tree, the solver state (containing all the domains, the
full constraint store and the propagation queue), and the \emph{user
time} spent since the beginning of the execution.
The specific attributes depend on the port.  
For example, the specific attributes for port \newVariable{} are the variable
identifier and its initial domain.
For the ports related to the search-tree the only specific attribute
is the node label. Specific attributes for other ports are described
in~\cite{langevine04}.

\begin{figure}[t,t,t]
\begin{alltt}\tt
 1 newVariable   v1=[0-268435455]
 2 newVariable   v2=[0-268435455]
 3 newConstraint c1 fd_element([v1,[2,5,7],v2])
 4 reduce c1 v1=[1,2,3] delta=[0,4-268435455]
 5 reduce c1 v2=[2,5,7] delta=[0-1,3-4,6,8-268435455]
 6 suspend c1
 7 newConstraint c4 x_eq_y([v2,v1])
 8 reduce   c4   v2 =[2]       delta=[5,7]
 9 reduce   c4   v1 =[2]       delta=[1,3]
10 suspend  c4
11 awake    c1
12 reject   c1
\end{alltt}
\vspace{-0.5cm}

\caption{A portion of trace}
\label{fig:analyse:trace_example}
\end{figure}

Fig.~\ref{fig:analyse:trace_example} presents the beginning of a trace
of a toy program in order to illustrate the events described above.
This program, 
\\{\tt fd\_element(I, [2,5,7],A), (A\#=I ; A\#=2)},
specifies that {\tt A} is a finite domain variable which is in
$\{2,5,7\}$ and {\tt I} is the index of the value of {\tt A} in this
list; moreover {\tt A} is either equal to {\tt I} or equal to 2. The
second alternative is the only feasible one.
The trace can be read as follows. 
The first two events are related to the introduction of two variables
{\tt\small v}$1$ and {\tt\small v}$2$, corresponding respectively
to {\tt I} and {\tt A}.  In Gnu-Prolog, variables are always created
with the maximum domain (from 0 to $2^{28}-1$
).
Then the first constraint is created: {\tt fd\_element} (event \#3).
This constraint makes two domain reductions (events \#4 and \#5): the
values removed from the domain of the first variable ({\tt I}) are
listed in {\tt delta}, the domain becomes $\{1,2,3\}$ and the domain
of {\tt A} becomes $\{2,5,7\}$, the only consistent values so far.
After these reductions, the constraint is suspended (event \#6).
The next constraint, {\tt A\#=I}, is added (event \#7).
Two reductions are done on variables {\tt A} and {\tt I}, the only
possible value for {\tt A} and {\tt I} to be equal is {\tt 2} (events
\#8 and \#9).
After these reductions, the constraint is suspended (event \#10).
The first constraint is awoken (event \#11).
If {\tt A} and {\tt I} are both equal to {\tt 2}, {\tt I} cannot be
the rank of {\tt A}. Indeed, the rank of {\tt 2} is {\tt 1} and the
value at rank {\tt 2} is {\tt 5}.  The constraint is therefore
rejected (event \#12).
The execution continues and finds the solution ({\tt A=2}, {\tt I=1})
This requires 20 other events not shown here.

\subsection{Patterns}
\label{sec:analyse:pattern_grammar}

\begin{figure}[t,t,t]
\begin{small}
\begin{tabular}{p{1.1cm}p{0.1cm}l}
\nonterminal{pattern} &::= &\terminal{\em label}\terminal{:} \terminal{when}
\nonterminal{evt\_pattern} ~\nonterminal{op\_synchro}~
\nonterminal{action\_list}\\[0.1cm] 
\nonterminal{op\_synchro}&::=& \terminal{do}~ |~ \terminal{do\_synchro}\\[0.1cm]
\nonterminal{action\_list}&::=& \nonterminal{action}~ \terminal{,}~
\nonterminal{action\_list}~ |~ \nonterminal{action}\\[0.1cm]
\nonterminal{action}&::=&
\terminal{current(}\nonterminal{list\_of\_attributes}\terminal{)}~ 
|~ \terminal{call(}\nonterminal{procedure}\terminal{)}\\[0.1cm]
\\
\nonterminal{evt\_pattern} &::= &\nonterminal{evt\_pattern}~ \terminal{or}~ \nonterminal{evt\_pattern}\hfill{\bf(1)}\quad\quad\\
&&| \nonterminal{evt\_pattern}~ \terminal{and}~ \nonterminal{evt\_pattern}\hfill{\bf(2)}\quad\quad\\
&&| \terminal{not}~ \nonterminal{evt\_pattern}\hfill{\bf(3)}\quad\quad\\
&&| \terminal{(} \nonterminal{evt\_pattern} \terminal{)}\hfill{\bf(4)}\quad\quad\\
&&| \nonterminal{condition}\hfill{\bf(5)}\quad\quad\\[0.1cm]
\nonterminal{condition}&::=& \terminal{{\em attribute}}~
\nonterminal{op2}~ \terminal{{\em value}}~ 
|~ \nonterminal{op1}\terminal{(}\terminal{{\em attribute}}\terminal{)}~~|~~
\terminal{true}\\[0.1cm] 
\nonterminal{op2}&::=& \terminal{<} | \terminal{>}
| \terminal{=} | \terminal{$\backslash$=} | \terminal{>=} | \terminal{=<}
| \terminal{in} | \terminal{notin} 
\\&&
| \terminal{contains} | \terminal{notcontains}\\[0.1cm]
\nonterminal{op1}&::=& \terminal{isNamed}
\end{tabular}
\end{small}
\caption{Grammar of event patterns}
\label{fig:analyse:grammar_patterns}
\end{figure}

We use patterns similar to the path rules of Bruegge and
Hibbard~\cite{bruegge83}.
Fig.~\ref{fig:analyse:grammar_patterns} presents the grammar of
patterns. A pattern contains four parts: a label, an event pattern, a
``synchronization'' operator and a list of actions.
An event pattern is a composition of elementary conditions using logical
conjunction, disjunction and negation. It specifies a class of execution
event. 
A ``synchronization'' operator tells whether the pattern is asynchronous
(\terminal{do}) or synchronous (\terminal{do\_synchro}).
An action specifies either to ask the tracer driver to collect
attribute values
(\terminal{current(}\nonterminal{list\_of\_attributes}\terminal{)}),
or to ask the analyzer to call a procedure
\terminal{call(}\nonterminal{procedure}\terminal{)}. Note that the
procedure is written in a language that the analyzer is able to
execute. This language is independent of the tracer driver.
An elementary condition concerns an attribute of the current event.

There are several kinds of attributes. Each kind has a specific set of
operators to build elementary conditions.
For example, most of the common attributes are integer (chrono, depth,
node label).  Classical operators can be used with those attributes:
equality, disequality ($\neq$), inequalities ($<$,
$\leq$, $>$ and $\geq$).
The \emph{port} attribute is the type of the current event. It has a
small set of possible values. The following operators can be used with
the port attribute: equality and disequality ($=$ and $\neq$) and two
set operators, \terminal{in} and
\terminal{notin}.
Constraint solvers manipulate a lot of constraints and variables. Often,
a trace analysis is only interested in a small subset of them. Operators
\terminal{in} and \terminal{notin}, applied to identifiers of entities
or name of the variables, can specify such subsets. Operators
\terminal{contains} and \terminal{notcontains} are used to express
conditions on domains. This set of operators is specific to the type
of execution we trace. It could be extended to cope with other types of
attributes.

\subsection{Examples of Patterns}
\label{sec:analyse:pattern_example}

\begin{figure}[t,t,t]
\begin{alltt}
visu_tree: 
  when port in [choicePoint, backTo, solution, failure]
  do current(port=P and node=N and depth=D and usertime=Time),
     call search_tree(P,N,D,Time)\vspace{0.1cm}
visu_cstr: 
  when port = post
  do current(cstr=C and cstrRep=Rep 
               and varC(cstr)=VarC),
     call new_cstr(C, Rep, VarC)\vspace{0.1cm}
visu_prop: 
  when port = reduce and isNamed(var) 
         and (not cstrType='assign')
         and delta notcontains [maxInt]
  do current(cstr=C and var=V),
  call spy_propag(C,V)\vspace{0.1cm}
leaf:      
  when port in [solution, failure]
  do_synchro current(port=P and node=N and depth=D),
             call new_leaf(P,N,D)\vspace{0.1cm}
symbolic:  
  when port in [reduce,suspend]
       and (cstrType = 'fd_element_var' 
           or cstrType = 'fd_exactly')
  do_synchro call symbolic_monitor
\end{alltt}
\caption{Examples of  patterns for visualization and monitoring}
\label{fig:analyse:patterns}
\end{figure}

\begin{figure}[t,t,t]
\begin{center}
\begin{minipage}[c]{0.45\linewidth}\includegraphics[angle=270,width=\linewidth]{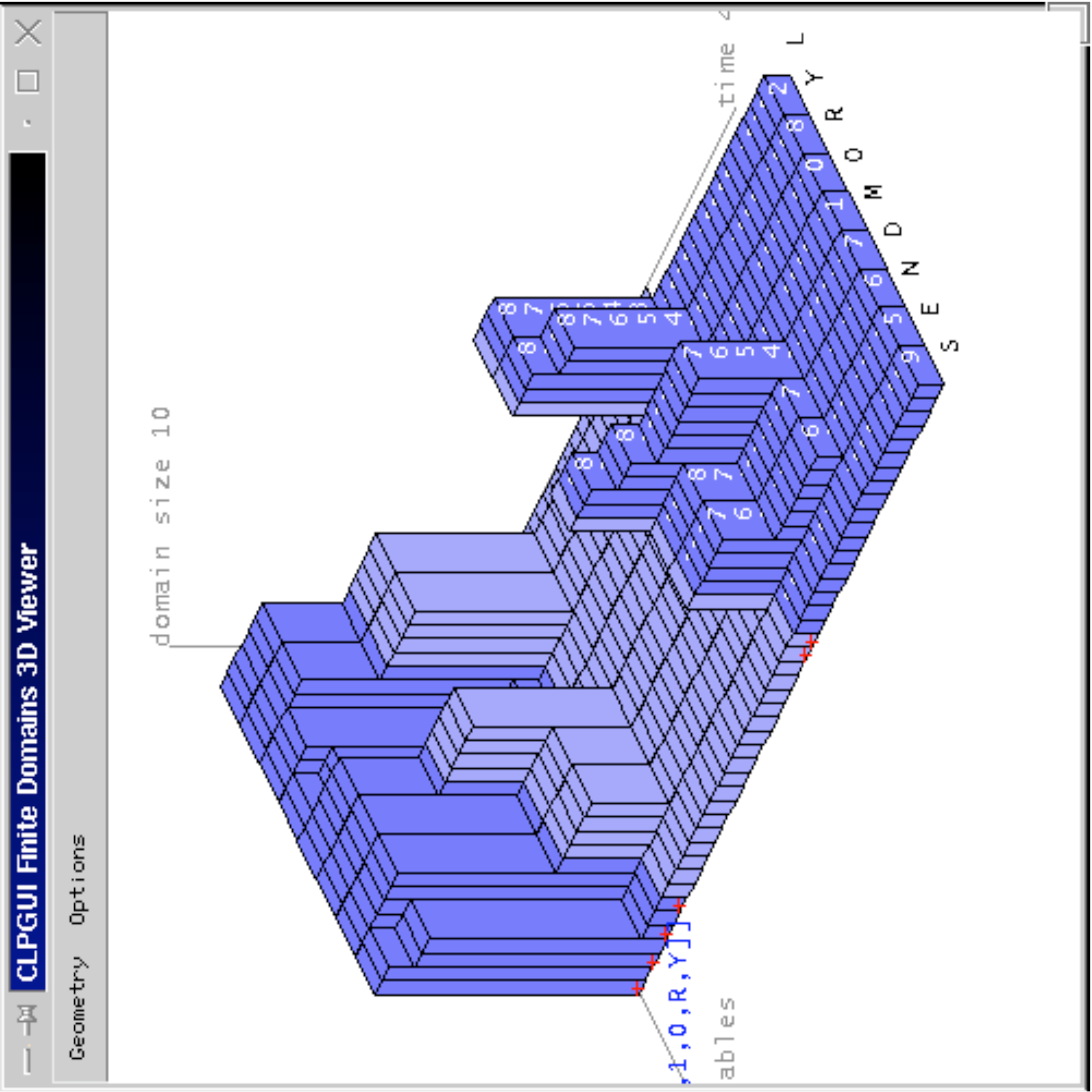}\end{minipage}\hfill
\begin{minipage}[c]{0.45\linewidth}\includegraphics[width=\linewidth]{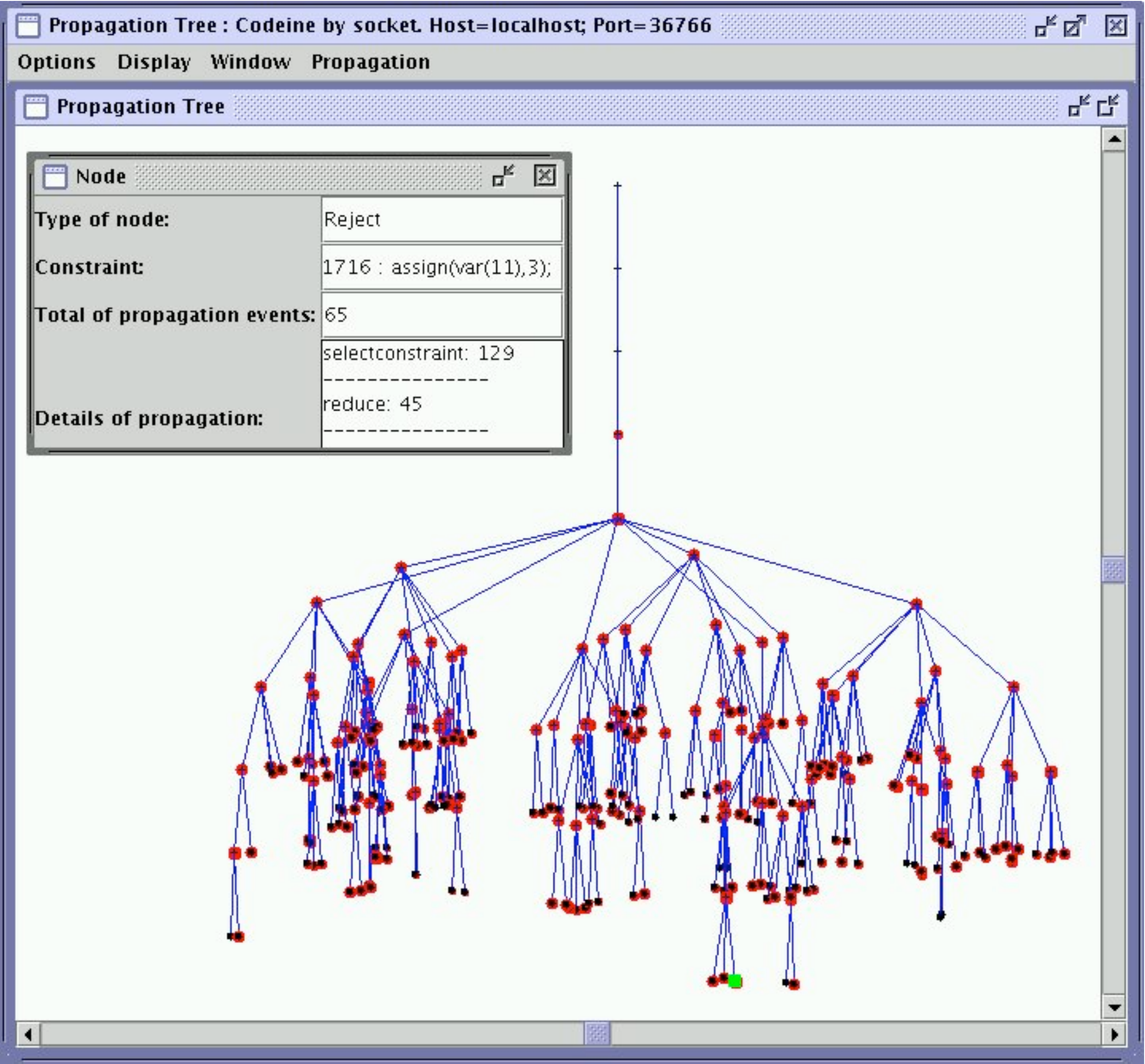}\end{minipage}
\caption{Two trace-based views based on the pattern {\tt visu\_cstr} of Figure~\ref{fig:analyse:patterns}.
}\label{fig:analyse:pavot}
\end{center}
\end{figure}

Fig.~\ref{fig:analyse:patterns} presents five patterns that can be
activated simultaneously.  The first three patterns are visualization
oriented: the first one, {\tt visu\_tree}, aims at constructing the
search-tree (creation of a choice-point, backtracking, failure and
solution are the relevant ports) with few details about each node: the
label of the node, its depth and the chronological time stamp of the
event. This enables the display of nodes to be weighted by the time
spent in their sub-tree.

The second pattern, {\tt visu\_cstr}, requests the trace of each
constraint-posting with the identifier of the constraint (a unique
integer), its representation (the name of the constraint with its
parameters) and the list of the involved variables (\emph{varC}).
Fig.~\ref{fig:analyse:pavot} gives two screen-shots of visualizations
that are built using pattern {\tt visu\_cstr}. The first
  picture is generated by CLPGUI~\cite{fages02}. The second picture is
  generated by Pavot, a graphical tool developed at INRIA
  Rocquencourt and connected to GNU-Prolog tracer
  driver \footnote{http://contraintes.inria.fr/\~{}arnaud/pavot/}.

The third pattern, {\tt visu\_prop}, requests the trace of all the
domain reductions made by constraints that do not come from the
assignment procedure and that do not remove the maximal integer value.
It stores the reducing constraint and the reduced variable. Those
data can be used to compute some statistics and to visualize the
impact of each constraint on its variables. Those three
patterns are asynchronous: the requested data are sufficient for the
visualization and the patterns do not have to be modified.

The fourth pattern, {\tt leaf}, synchronizes the execution at each
leaf of the search-tree (solution or failure). At those events, the
{\tt new\_leaf} function can interact with the tracer to investigate
the execution state.

The last pattern, {\tt symbolic}, is more monitoring-oriented: it
freezes the execution at each domain reduction made by a symbolic
constraint such as \emph{element} (on variables) or
\emph{exactly}. This pattern allows the monitoring of the filtering
algorithms used for these two constraints.

\section{Analyzer Mediator}
\label{sec:analyse:request} 

The analyzer mediator is the interface between the tracer driver and
the analyzers. It specifies to the tracer driver what events are
needed and may execute specific actions for each class of relevant
events. The mediator can supervise several analyses at a time.  Each
analysis has its own purpose and uses specific pieces of trace
data. The independence of the concurrent analyses is ensured by the
mediator that centralizes the communication with the tracer driver and
distributes the trace data to the ongoing analyses. The advantage is
that if a piece of trace information is needed by several analyses, it
is sent over the interface only once.

When a synchronous event has been sent to the mediator, the requests
that an analyzer can send to the driver are of three kinds.
Firstly, the analyzer can ask for additional data about the current
event. 
Secondly, the analyzer can modify the set of active event patterns, to
be checked by the tracer driver. 
Thirdly, the analyzer can notify the 
tracer driver that the execution can be resumed.
The actual requests are as follows.

\begin{description}
\item
{{\tt current}} specifies a list of event attributes to
retrieve in the current execution event.
The tracer retrieves the requested pieces of data. It sends the data
as a list of pairs $(attribute, value)$.\quad
\item
{{\tt reset}} deletes all the active event patterns and their
labels. 
\item
{{\tt remove}} deletes the active patterns whose labels are
specified in the parameter.
\item
{{\tt add}} inserts in the active patterns, the event patterns
  specified in the parameter, following the grammar described in
  Figure~\ref{fig:analyse:grammar_patterns}.
\item {{\tt go}} notifies the tracer driver that the traced execution
  is to be resumed.

\end{description}

\begin{figure}[t,t,t]
\begin{alltt}
step :-
  {\bf{}reset},
  {\bf{}add}([step:when true dosynchro call(tracer_toplevel)]),
  {\bf{}go}.

skip_reductions :-
  {\bf{}current}(cstr = CId and port = P),
  {\bf{}reset},
  (  P == awake
  -> {\bf{}add}([sr:when cstr = CId and port in [suspend,reject,entail] 
                    dosynchro call(tracer_toplevel)]),
  ;  {\bf{}add}([step:when true do_synchro call(tracer_toplevel)])
  ),
  {\bf{}go}.
\end{alltt}
\caption{Implementation of two tracing commands}
\label{fig:analyse:standard_tracing}
\end{figure}

Fig.~\ref{fig:analyse:standard_tracing} illustrates the use of the
primitives to implement two tracing commands. Let us assume that the
analyzer is a (possibly simplified) Prolog interpreter, as for example
in Opium~\cite{ducasse99}.
Command {\tt step} enables execution to go to the very next event. It simply
resets all patterns and adds one that will match any event and call,
with synchronous interactions, the tracer toplevel.
Command {\tt skip\_reductions} enables execution to skip the
details of variable domain reductions when encountering the awakening of a
constraint. It first checks the current port. If it is \awake{} it
asks to go to the suspension of this constraint. There, the user will,
for example, be able to check the value of the domains after all the
reductions. If the command is called on an event of another type, it
simply acts as {\tt step}.

\section{Filtering Mechanism}
\label{sec:analyse:filtering} 

This section describes in detail the critical issue of the filtering
mechanism.  At each execution event, it is called to test the
relevance of the event with respect to the active patterns. Notice
that the execution of a program with constraints can lead to several
millions of execution events per second. Therefore, the efficiency of
the event filtering is a key issue.

In the following, we first describe the algorithm of the tracer
driver.  Then we specify the automata which drive the matching of
events against active patterns.  We discuss some specialisation
issues.  We give some details about the incremental handling of
patterns.  Lastly, we emphasize that event attributes are computed
only upon demand.

\subsection{Tracer Driver Algorithm}
\label{pattern:evaluation:section}

\begin{figure}[t,t,t]
\begin{alltt}
 1.  proc tracerDriver(\(P:\) set of active patterns)
 2.    \(tagged \gets \emptyset\)
 3.    for each {\em{}p} \(\in P\) do
 4.     if match({\em{}p}) then \(tagged \gets tagged \cup \{p\}\)
 5.    end for
 6.   T \(\gets \{requested\_data(p) | p\in\mathrm{tagged}\}\)
 7.   send_trace_data(T, \(label(tagged)\))
 8.   if \(synchronous(tagged)  \neq \emptyset\) then
 9.     notify(\(synchronous(tagged)\))
10.     repeat
11.        \(request\) \(\gets\) receive_from analyzer()
12.        execute(\(request\))
13.     until \(request\) = \(\mathbf{go}\)
14.    end if
15.  end proc
\end{alltt}
\vspace{-0.5cm}
\caption{Algorithm of the Tracer Driver}
\label{fig:analyse:tracer_driver_algo}
\end{figure}

When an execution event occurs, the tracer is called. The tracer
collects some data to maintain its own data structures and then calls
the tracer driver. The algorithm of the tracer driver is given in
Fig.~\ref{fig:analyse:tracer_driver_algo}.  The filtering mechanism
can handle several active event patterns.
For each pattern, if the current event matches the pattern the latter
is tagged as activated;  whatever the matching result, the next
pattern is checked (lines 3-5).
When no more patterns have to be checked, the tagged patterns are
processed; the union of requested pieces of data is sent as trace data
with the labels of the tagged patterns (lines 6-7). 
If at least one synchronous pattern is tagged, a signal is sent to the
analyzer; the tracer driver waits for requests coming from the
analyzer and processes them until the \(\mathbf{go}\) primitive is
sent by the analyzer (lines8-14).

\subsection{Pattern Automata}
\label{automata:section}

The matching of an event against a pattern is driven by an automaton
where each state is labeled by an elementary condition with two
possible transitions: true or false. The automaton has two final
states, $\mathbf{true}$ and $\mathbf{false}$. If the $\mathbf{true}$
state is reached, the event is said to match the pattern.
Each automaton results from the compilation of an event pattern. This
compilation is  inspired by the  evaluation of Boolean
expressions in imperative languages.  It has been proven that it
minimizes the number of conditions to check~\cite{wilhelm95}. 
Examples of automata are given in Figure~\ref{graphs:figure},
section~\ref{automata:examples:section}.

\subsection{Specialization According to the Port}
\label{optimization:section}

As seen in Section~\ref{sec:analyse:pattern}, the port is a special
attribute since it denotes the \emph{type} of the execution event
being traced.
A port corresponds to specific parts of the solver code where a call
to a function of the tracer has been hooked.  For example, in the
tracer we use, there are four hooks for the \reduce{} port, embedded into
four specific functions that make domain reductions in four different ways.
Furthermore, specific attributes depend on the port.
As a consequence, the port is central in the pattern
specification. For most patterns, a condition on the port will be
explicit.
When an event occurs, it is useless to call the tracer driver if no
pattern is relevant for the port of this event. Therefore, for each
port, a flag in the related hooks indicates whether the port appears
in at least one pattern.  This simple mechanism avoids useless calls
to the tracer driver.

\subsection{Examples of Pattern Automata}
\label{automata:examples:section}

\begin{figure}[t,t,t]
\includegraphics[width=\linewidth]{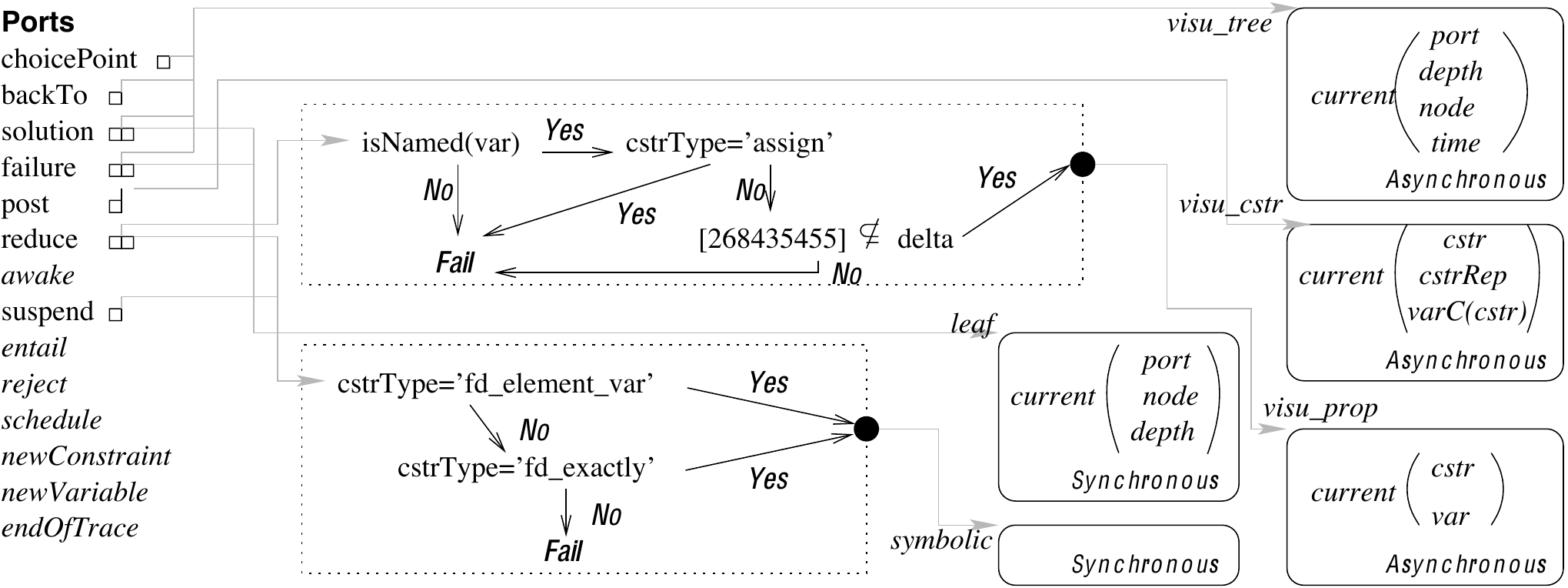}

\caption{Internal representation of the five patterns of Figure~\ref{fig:analyse:patterns}}
\label{graphs:figure}
\end{figure}

Fig.~\ref{graphs:figure} shows the internal representation of the five
patterns presented in Figure~\ref{fig:analyse:patterns},
Section~\ref{sec:analyse:pattern_example}.  The 14 ports are
represented on the left-hand side. The irrelevant ports are in
italic. The relevant ports are linked to their corresponding
patterns. Only two automata are necessary since three of the patterns
check the port only. A set of actions is assigned to each
automaton. This set is attached to ``synchronous'' or ``asynchronous''
and the label of the pattern.

\subsection{Incremental Pattern Management}
\label{incremental:management:section}

Since each active pattern is a specific automaton (or a list of
specific automata when split), the {\tt add} primitive has just to
compile the new $n$ patterns into $m$ automata, linked
them with their respective ports and store the labels with the lists
of resulting automata.  The {\tt remove} primitive has just to delete
the automata associated to the specified labels and to erase the dead
links. After each operation, the port-filtering flags are updated so
as to take into account the new state of the active patterns.

\section{Prototype Implementation}
\label{tracer:implementation:section}

In this Section, we briefly present the prototype implementation. 
In particular, in order for the overall architecture to be efficient,
it is essential that the tracer is lazy.

Trace information must not be computed if it is not
explicitly required by a pattern.
Indeed, an execution has many events and events potentially have many
attributes.  Most of them are not straightforwardly available, they
have to be computed from the execution state or from the debugging
data of the tracer.  Systematically computing all the attributes at
all the execution events would be terribly inefficient.

Fortunately, not all the attributes need to be computed at each event.
According to the active patterns, only a subset of the attributes is
needed: firstly, the attributes necessary to check the relevance of
the current event with respect to the patterns, and secondly, the
attributes requested by the patterns in case of matching.
Therefore, the tracer must not compute any trace attribute before it
is needed. When a specific attribute is needed, it is computed and its
value is stored until the end of the checking of the current event. If
an attribute is used in several conditions, it is computed only once.

The tracer implemented in the current prototype, Codeine, strictly
follows this guideline. Some core tracer mechanisms are needed to
handle the debugging information (see~\cite{langevine03} for more
details). As shown in Sec.~\ref{sec:analyse:performance}, the overhead
induced by these mechanisms is marginal, even though constraint
solvers do manipulate large and complex data.

Currently, Codeine is implemented in 6800 lines of C including
comments.  The tracer driver, including the communication mechanisms,
is 1700 lines of C. 
The codeine tracer including the tracer driver is available under GNU
Public Licence. The set of debugging and visualization tools called
Pavot has been developed by Arnaud Guillaume and Ludovic
Langevine. Both systems are available on line\footnote{ They can be
retrieved at http://contraintes.inria.fr/\~{}langevin/codeine/ and
http://contraintes.inria.fr/\~{}arnaud/pavot/.}.

\section{Experimental Results}
\label{sec:analyse:performance}

This section assesses the performance of the tracer driver and its
effects on the cost of the trace generation and communication.  
Sections~\ref{sec:performance:methodology}
and~\ref{sec:performance:setting} describe the methodology of the
experiments and the experimental setting.
Section~\ref{sec:performance:bench} lists the benchmark programs.
Sections~\ref{sec:performance:tracer:overhead},
\ref{sec:performance:driver:overhead}
and~\ref{sec:performance:comm:overhead} respectively discuss the
tracer overhead, the tracer driver overhead and the communication
overhead.

\subsection{Methodology of the Experiments}
\label{sec:performance:methodology}

When tracing a program, some time is spent in the program execution
($T_{prog}$), some time is spent in the core mechanisms of the tracer
($\Delta_{tracer}$), some time is spent in the tracer driver
($\Delta_{driver}$), some time is spent generating the requested trace
and sending it to the analysis process ($\Delta_{gcom}$), and lastly some
time is spent in the analyses ($\Delta_{ana}$). Hence, if we call $T$
the execution time of a traced and analyzed program, we
approximately have:

\begin{center}
$T \simeq T_{prog} + \Delta_{tracer} + \Delta_{driver} + \Delta_{gcom} + \Delta_{ana}$.
\end{center}

The mediator is a simple switch. The time taken by its execution is
negligible compared to the time taken by the simplest analysis,
namely the display of trace information. Trace analysis takes a time
which can vary considerably according to the nature of the analysis. The
focus of this article is not to discuss which analyses can be achieved
in reasonable time but to show that a flexible analysis environment
can be offered at a low overhead. Therefore, in the following
measurements $\Delta_{ana}= 0$.

\subsection{Experimental Setting}
\label{sec:performance:setting}

The experiments have been run on a PC, with a 2.4\,GHz
Pentium {\sc iv}, 512 Kb of cache, 1~Gb of RAM, running under the
GNU/Linux~2.4.18 operating system.
We used the most recent stable release of \gprolog{}  (1.2.16).  The
tracer is an instrumentation of the source code of this 
version and has been compiled 
by {\tt gcc-2.95.4}.
The execution times have been measured with the \gprolog{} profiling
facility whose accuracy is said to be 1\,ms. The measured executions
consist of a batch of executions such that each measured time is at
least 20 seconds.
The measured time is the sum of \emph{system} and \emph{user} times.
Each experimental time given below is the average time of a series of
ten measurements.  In each series, the maximal relative deviation was
smaller than 1\,\%.

\subsection{Benchmark Programs}
\label{sec:performance:bench}

The 9 benchmark programs\footnote{Their source code is available at
http://contraintes.inria.fr/\~{}langevin/codeine/benchmarks} are
listed in Table~\ref{tab:analyse:characteristic_programs}, sorted by
increasing number of trace events.
  Magic(100), square(4), golomb(8) and golfer(5,4,4) are part of
  CSPLib, a benchmark library for constraints by
  Gent~and~Walsh~\cite{csplib}.
The golomb(8) program is executed with two strategies which exhibit
very different response times.
Those four programs have been chosen for their significant execution
time and for the variety of constraints they involve.
Four other programs have been added to cover more specific aspects of
the solver mechanisms:
Pascal Van Hentenryck's bridge problem, implementation  of~\cite{gnuprolog};
two instances of the $n$-queens problem; and
``propag'', which proves the infeasibility of 
\begin{eqnarray}
1 \leq x \wedge y \leq 70000000 \wedge x < y \wedge y < x.  \nonumber
\end{eqnarray}
The interest of the latter is the long stage of propagation involving
one of the simplest and the most optimized constraints \gprolog{}
provides: bound consistency for a strict inequality. Therefore, this
program  the worst kind of  case for the propagation instrumentation.

\begin{table}
\begin{center}
\begin{tabular}{|lrrrrrr|}
\hline
{\bf Program}& \#events&Trace&$T_{prog}$&$\varepsilon$&$R_{tr.}$&Dev.\\
             & ($10^6$)&Size&$(ms)$    &$(ns)$  &         &for $T_x$\\
             &         &(Gb)&        &        &         &in \%\\
\hline
\emph{bridge}        & 0.2   & 0.1  & 14     & 72  & 1.21 & $\leq 0.4$\\
\emph{queens(256)}   & 0.8   & 1.5  & 173    & 210 & 1.14 & $\leq 0.2$\\
\emph{magic(100)}    & 3.2   & 1.4  & 215    & 66  & 1.03 & $\leq 0.2$\\
\emph{square(24)}    & 4.2   & 20.8 & 372    & 88  & 1.05 & $\leq 0.6$\\
\emph{golombF}       & 15.5  & 3.4  & 7,201  & 464 & 1.01 & $\leq 0.4$\\
\emph{golomb}        & 38.4  & 7.9  & 1,721  & 45  & 1.00 & $\leq 0.5$\\
\emph{golfer(5,4,4)} & 61.0  & >30  & 3,255  & 53  & 1.05 & $\leq 0.7$\\
\emph{propag}        & 280.0 & >30  & 3,813  & 14  & 1.28 & $\leq 1.0$\\
\emph{queens(14)}    & 394.5 & >30  & 17,060 & 43  & 1.08 & $\leq 0.4$\\
\hline
\end{tabular}
\end{center}
\caption{Benchmark Programs and tracer overhead}
\label{tab:analyse:characteristic_programs}
\end{table}

The benchmark programs have executions large enough for the
measurements to be meaningful. They range from 
200,000 events to about 400 millions events. 
Furthermore, they represent a wide range of CLP(FD) programs.

The third column gives  the size of the traces of the benchmarked
programs for the default trace model.
All executions but the smallest one exhibit more than a gigabyte, for
executions sometimes less than a second. It is therefore not
feasible to systematically generate such an amount of
information. As a matter of fact measuring these sizes took us 
hours and, in the last three cases, exhausted our patience!
Note that the size of the trace is not strictly proportional to the
number of events because different sets of attributes are collected at
each type of events. For example, for domain reductions, several
attributes about variables, constraints and domains are collected
while other types of events simply collect the name of the
corresponding constraint.

The fourth column gives $T_{prog}$, the execution time in $ms$ of the program
simply run by \gprolog{}.
The fifth column shows the average time of execution per event
$\varepsilon= \frac{T_{prog}}{\mathrm{\#events}}$. It is between 14 ns
and 464 ns per event. For most of the suite $\varepsilon$ is around
50ns. The three notable exceptions are \emph{propag} ($\varepsilon =
14$ ns), \emph{queens(256)} ($\varepsilon = 210$ ns) and \emph{golombF}
($\varepsilon = 464$ ns). The low $\varepsilon$ is due to the efficiency
of the propagation stage for the constraints involved in this
computation. The large $\varepsilon$s are due to a lower proportion of
``fine-grained'' events.

\subsection{Tracer Overhead}
\label{sec:performance:tracer:overhead}

The sixth column of Table~\ref{tab:analyse:characteristic_programs}
also gives the results of the measurements of the overhead of the core
tracer mechanisms, $R_{tr(acer)}$, which is defined as the ratio:
\begin{center}
$R_{tr(acer)} = \frac{T_{tracer}}{T_{prog}}$. 
\end{center}
where the measure of 
\begin{center}
$T_{tracer} \simeq T_{prog} + \Delta_{tracer}$ 
\end{center}
is the execution time of the program run by the tracer without any
pattern activated. The tracer maintains its own data for all
events. However, no attribute is calculated and no trace is generated.

The  seventh column gives the maximum deviation for $T_{prog}$ and
$T_{tracer}$.

\paragraph{Core tracer mechanisms can be permanently activated.\quad}

For all the measured executions $R_{tracer}$ is less than 30\% in the
worst case, and less than 5\% for five traced programs.
The results for $R_{tracer}$ are very positive; they mean that the
core mechanisms of the tracer can be systematically activated. Users
will hardly notice the overhead. Therefore, while developing
programs, users can directly work in ``traced'' mode; they do not need
to switch from untraced to traced environments. This is a great
comfort.
As soon as they need to trace they can immediately get
information.

\subsection{Tracer Driver Overhead}
\label{sec:performance:driver:overhead}

The measure of 
$T_{driver}\simeq T_{prog} + \Delta_{trace}+ \Delta_{driver}$ 
is the execution time of the program run by the tracer with the
filtering procedure activated for generic patterns. Only the
attributes necessary for the requested patterns are calculated at
relevant events.  In order for $\Delta_{gcom}$ to be zero, the
patterns are designed such that no event matches them. One run is done
per pattern. The patterns are listed in
Figure~\ref{tab:analyse:patterns:driver}. Pattern {\bf 1a} is checked
on few events and on one costly attribute only. 
Pattern {\bf 2a} is checked on two costly attributes and on numerous
events.  Indeed, \reduce{} events trace the main mechanism of the
propagation and they are significantly more numerous than the other
types of events.
Pattern {\bf
  3a} is checked on all events and on one cheap attribute. Pattern
{\bf 4a} is checked on all events and systematically on three
attributes.

\begin{figure}
\begin{quote}
{\em {\bf 1a.} when port=post and isNamed(cname) 
\\~~~~~do current(port,chrono,cident)}.
\\~~

{\em  {\bf 2a.} when port=reduce and 
\\~~~~~~~~~~~(isNamed(vname) and isNamed(cname)) 
\\~~~~~do current(port,chrono,cident)}.
\\~~

{\em {\bf 3a.} when chrono=0 do current(chrono)}. 
\\~~

{\em {\bf 4a.} when depth=50000 or (chrono>=1 and node=9999999) 
\\~~~~~do current(chrono,depth)}.
\\~~
 
{\em {\bf 5a}}: patterns 1a, 2a, 3a and 4a activated simultaneously.
\end{quote}

\caption{Patterns used to measure the tracer driver overhead}
\label{tab:analyse:patterns:driver}
\end{figure}

\begin{figure}[tttt]
\begin{center}
\includegraphics[width=\linewidth]{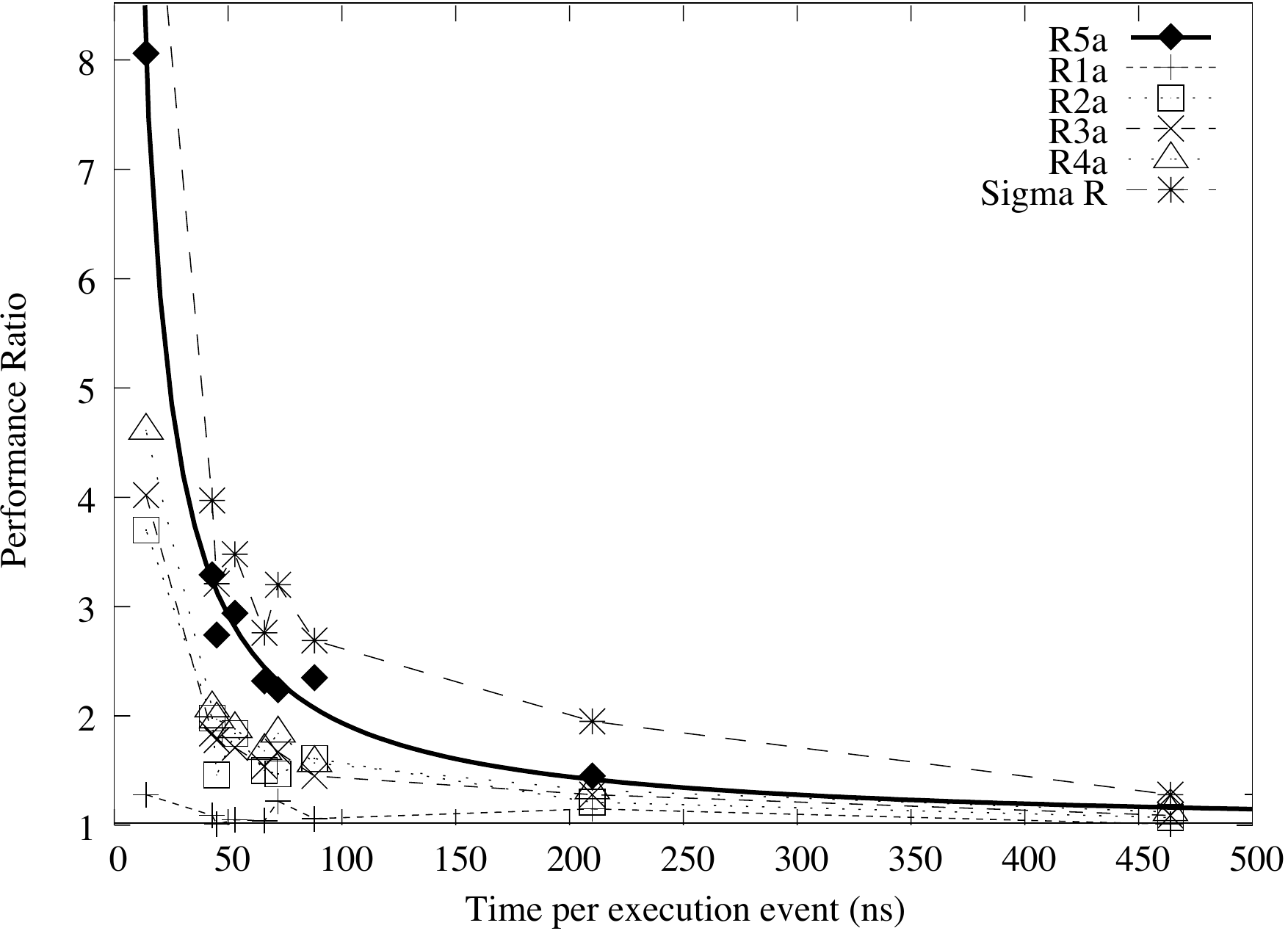}
\caption{Cost of the tracer driver mechanisms for patterns 1a to 5a compared to
$\mathbf{\varepsilon}$ }
\label{tab:analyse:perf_bench}
\end{center}
\end{figure}

\begin{figure}[tttt]
\begin{center}
\includegraphics[width=\linewidth]{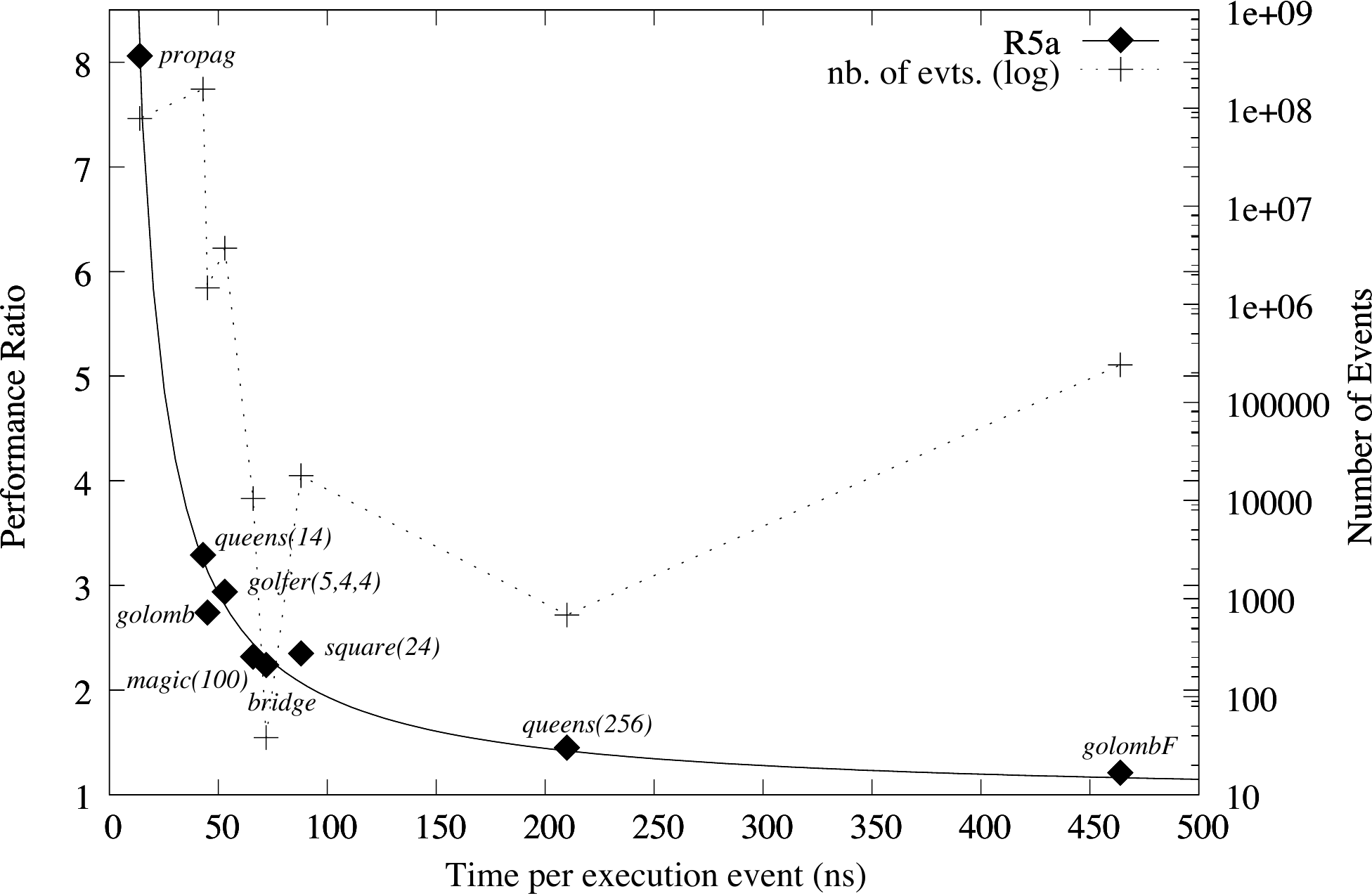}
\caption{Cost of the tracer driver mechanisms compared to
the number of events}
\label{tab:analyse:perf:nbevt}
\end{center}
\end{figure}

In order to measure the overhead of the tracer driver, for each of the 5
patterns a ratio $R_{pattern}$ is computed for all benchmark programs :
\begin{center}
$R_{pattern} = \frac{T_{driver}}{T_{prog}} = 1+\frac{\Delta_{driver}^{pattern}}{T_{prog}}$. 
\end{center}

Figure~\ref{tab:analyse:perf_bench} displays all the ratios compared
to the average time per event ($\varepsilon$) of the programs. The
smallest value of $\varepsilon$, 14, corresponds to program \emph{propag} and the
biggest value, 464, corresponds to program golombF.

In addition, the figure shows the curve 
$Sigma \;R = R_{1a}+R_{2a}+R_{3a}+R_{4a}-3$, that
adds the overheads of the four separated patterns.

\paragraph{Tracer driver overhead is acceptable.\quad}

In Figure~\ref{tab:analyse:perf_bench}, for all but one program,
$R_{pattern}$ is negligible for the very simple patterns and less than
3.5 for pattern {\bf 5a} which is the combination of the other four
patterns. For programs with a large $\varepsilon$, even searching for
pattern {\bf 5a} is negligible. In the worst case, for \emph{propag},
an overhead ratio of 8 is still acceptable.

\paragraph{No overhead for simultaneous search for patterns.\quad}
When $n$ patterns are checked simultaneously they already save 
$(n-1)T_{tracer}$ compared to the search in sequence which requires
the program to be executed $n$ times instead of one time.
Figure~\ref{tab:analyse:perf_bench} further shows that the curve
$Sigma \;R = R_{1a}+R_{2a}+R_{3a}+R_{4a}-3$, is above the curve of
$R_{5a}$. 
Hence 
\begin{center}
$\Delta_{driver}^{1a} + \Delta_{driver}^{2a} + \Delta_{driver}^{3a}
+ \Delta_{driver}^{4a} > \Delta_{driver}^{(1|2|3|4)a}$. 
\end{center}
This means that not only is there no overhead in the filtering
mechanism induced by the simultaneous search, but there is even a minor
gain, due to the factorization in the automata described
Section~\ref{automata:section}.

\paragraph{Tracer driver overhead is predictable.\quad} 
The measured points of Figure~\ref{tab:analyse:perf_bench} can be
interpolated with curves of the form
\begin{center}
$R_{driver} = a + b/\varepsilon$.  
\end{center}
Figure~\ref{tab:analyse:perf:nbevt} recalls the curve for pattern {\bf
5a} and gives the number of events. It shows that there is no
correlation between the size of the trace and the tracer driver
overhead.

Those results mean that the tracer and tracer driver overheads per
event can be approximated to constants depending on the patterns and
these constants are {\em independent of the traced program.} Indeed,
let us assume that that $\Delta_{tracer}= N \delta_{tracer}$
and $\Delta_{driver}=N \delta_{driver}$ where $N$ is the number of
events of an execution, $\delta_{tracer}$ and $\delta_{driver}$
are the average time per event taken respectively by the core tracer
mechanism and the tracer driver, for all the programs. We have also
already assumed that
\begin{center}
$T_{driver}\simeq T_{prog} + \Delta_{tracer}+
\Delta_{driver}$,
\end{center}
and we have

\begin{center}
\begin{tabular}{c}
$R_{driver} = \frac{T_{driver}}{T_{prog}}$,~~ and~~
%
$T_{prog}  = N \varepsilon$\\\\
\end{tabular}
\end{center}
therefore
\begin{center}
\begin{tabular}{c}
$R_{driver}\simeq \frac{T_{prog} + \Delta_{tracer}+
\Delta_{driver}}{T_{prog}}$ \\\\
$R_{driver}\simeq  1+ \frac{N (\delta_{tracer}+ \delta_{driver})}{N \varepsilon}$\\\\
$R_{driver}\simeq  1+ \frac{\delta_{tracer}+ \delta_{driver}}{\varepsilon}$\\\\
\end{tabular}
\end{center}

The measured $R_{driver}$ for pattern {\bf 5a} is
\begin{center}
$R_{driver}= 0.95  + \frac{100ns}{\varepsilon}$.
\end{center}

For pattern {\bf 5a}, the average time per event taken by the core
tracer mechanism and the tracer driver ($\delta_{tracer}+
\delta_{driver}$) can therefore be approximated to $100ns$.

The $R_{driver}$ overhead could thus be made predictable.  For a given
program, it is easy to automatically measure $\varepsilon$, the
average time of execution per event. For a library of patterns
$\delta_{tracer}+ \delta_{driver}$ can be computed for each
pattern. We have shown above that the overhead of the simultaneous
search for different patterns can be over approximated by the sum of
all the overheads.
Our environment could therefore provide estimation mechanisms. When
$\varepsilon$ would be too small compared to $\delta_{tracer}+
\delta_{driver}$ the user would be warned that the overhead may become
large.

\subsection{Communication Overhead}
\label{sec:performance:comm:overhead}

The measure of 
\begin{center}
$T_{gcom} \simeq T_{tracer} + \Delta_{driver} + \Delta_{gcom}$ 
\end{center}
is the execution time of the program run by the tracer. 
A new set of patterns are used so that some events match the patterns,
the requested attributes of the matched events are generated and sent
to a degenerated version of the mediator: a C program that simply
reads the trace data on its standard input.
We show the result of program golomb(8) which has a median number of
events and has a median $\varepsilon$.

\begin{figure}
\begin{quote}
{\em {\bf 6b.} cstr: when port=post do
current(chrono,cident,cinternal).} 
\\~~~~~{\em tree: when port in [failure,backTo,
choicePoint,solution] do current(chrono,node,port).} 
\\~~

{\em {\bf 7b.} newvar: when port=newVariable do current(chrono, vident, vname).}
\\~~~~~{\em dom: when port in [choicePoint,backTo,solution] 
\\~~~~~~~~~~~~do current(chrono,node,port,named\_vars,full\_dom).}  
\\~~

{\em {\bf 8b.} propag1: when port=reduce do current(chrono).}
\\~~

{\em {\bf 9b.} propag2: when port=awake do current(chrono).}
\end{quote}

\caption{Event patterns used to assess the trace generation
  and the communication overhead}
\label{fig:analyse:patterns_com}
\end{figure}

The patterns are listed in
Figure~\ref{fig:analyse:patterns_com}.
Pattern {\bf 6b}, composed of two basic patterns, allows
a ``bare'' search tree to be constructed, as shown by most debugging tools. 
Pattern {\bf 7b} (two basic patterns) allows the display of 3D views of
variable updates as shown in Figure~\ref{fig:analyse:pavot}.
Pattern {\bf 8b} and
pattern {\bf 9b} provide two different execution details to decorate
search trees. Depending on the tool settings, three different visual
clues can be displayed. One is shown in
Figure~\ref{fig:analyse:pavot},
Section~\ref{sec:analyse:pattern_example}.

\begin{table}[tttt]
\begin{center}\footnotesize
\begin{tabular}{|c||r|r|r|r|r|}
    \hline
  \multicolumn{6}{|c|}{Program: golomb(8)\qquad $\varepsilon=45\mathrm{ns}$\qquad $T_{prog=1.73\mathrm{s}}$}\\
  \hline
    Patterns & Matched   & XML Trace     & Elapsed  & $R_{driver}$& $R_{gcom}$\\
             & events     &  size     & time & & \\
             & $(10^6)$ & (Mbytes) &  (s) & & \\
  \hline
6b         & 0.36  &  21 & 4.50 & 1.03 & 2.6\\
7b         & 0.13  & 111& 16.17 & 1.02 & 9.35\\
8b         & 5.04  & 141& 33.57  & 1.14 & 19.40\\
9b         & 14.58 & 394& 89.40  & 1.32 & 51.68\\
\hline
(6|7)b     & 0.36  & 124 & 17.47  & 1.04 & 10.09\\
(6|8)b     & 5.40  & 162 & 36.08  & 1.15 & 20.85\\
(6|9)b     & 14.94 & 415 & 92.71  & 1.33 & 53.59\\
(6|8|9)b   & 19.97 & 556 & 122.72  & 1.44 & 70.93\\
(6|7|8|9)b & 19.97 & 660 & 136.80 & 1.44& 79.07\\
\hline
\emph{default trace} 
           & 38.36 & 7,910  & 393.08& 1.96 & 227.21 \\
\hline
\end{tabular}
\caption{Cost of the trace generation and communication for program golomb(8)}
\label{tab:analyse:perf_bench_comm}
\end{center}
\end{table}

Table~\ref{tab:analyse:perf_bench_comm} gives the results for the
above patterns and some of their combinations. All combinations
correspond to existing tools. For example, combining {\bf 6b} with
{\bf 8b} or/and {\bf 4b} allows a Christmas tree as shown in
Figure~\ref{fig:analyse:pavot} to be constructed with two different
parametrization.
The 2\textsuperscript{nd} column gives the
number of events which match the pattern.
The 3\textsuperscript{rd} column gives the size of the resulting XML trace
as it is sent to the tool.
The 4\textsuperscript{th} column gives the elapsed time\footnote{Here
  system and user time are not sufficient because two processes are
  involved. $T_{prog}$ has been re-measured in the same conditions.}.
The 5\textsuperscript{th} column gives the ratio $R_{driver}$,
recomputed for each pattern.
The 6\textsuperscript{th} column gives the ratio $R_{gcom} =
\frac{T_{gcom}}{T_{prog}}$.

\paragraph{Filtered trace is more efficient and more accurate than
  default trace.\quad} The last line gives results for the
\emph{default} trace.  On the one hand, the \emph{default} trace
contains twice as many events as the trace generated by pattern {\bf
(6|7|8|9)b}; it also contains more attributes than requested by the
pattern; as a result, its size is ten times larger  and its
$R_{gcom}$ overhead is three times larger.
On the other hand, the \emph{default} trace does not contain all the
attributes. In that particular case, some relevant attributes are
missing in the default trace while they are present in the trace
generated by pattern {\bf (6|7|8|9)b}.  These attributes can be
reconstructed by the analysis module, but this requires further
computation and memory resources.

As a consequence, the tracer driver approach that we propose is more
efficient than sending over a default trace, even to construct
sophisticated graphical views. The accuracy and the lower volume of
the trace ease its post-processing by the debugging tools.

\paragraph{Answering queries is more efficient
  than displaying traces.\quad} 
$R_{gcom}$ is always much larger than $R_{driver}$, from $2.6$ to
$79.07$ in our example. Therefore, queries using patterns that
drastically filter the trace have significantly better response time
than queries that first display the trace before analyzing it.

When debugging, programmers often know what they want to check. In
that case they are able to specify queries that demand a simple
answer. In such a case our approach is significantly better than
systematically sending the whole trace information to an analyzer.

\paragraph{No need to restrict the trace information  a priori.\quad}
Many tracers restrict the trace information a priori in order to
reduce the volume of trace sent to an analyzer. This restricts the
possibilities of the dynamic analyses without preventing the
big size and time overhead as shown above with the default trace which
does not contain important information while being huge.

With our approach, trace information which is not requested does not
cost much, therefore our trace model can afford to be very rich. This
makes it easier to add new dynamic analyses.

\paragraph{Performance is comparable to the state-of-the-practice.\quad}
$R_{gcom}$ varies from 2.6 to 79.07. To give a comparison, the Mercury
tracer of Somogyi and Henderson~\cite{somogyi99} is regularly used by
Mercury developers. For executions of size equivalent to those of our
measurements, the Mercury tracer overhead has been measured from 2 to
15, with an average of 7~\cite{JD02tplp}.  Hence the ratios for
patterns {\bf 6b}, {\bf 7b} and {\bf 6|7b} are quite similar to the
state-of-the-practice debuggers. The other patterns show an overhead
that can discourage interactive usage. However, these patterns are
thought of more for monitoring than debugging when the interaction
does not have to be done in real time. Note, furthermore, that for the
measured programs, the absolute response time is still on the range of
two minutes for the worst case. When debugging, this is still
acceptable, especially if one considers that in the relevant cases the
alternatives to the uses of tools like ours are likely to be much
worse.

\paragraph{}
Our approach therefore allows to have the tracer present but idle by
default. When a problem is encountered, simple queries can be set to
localize roughly the source of the problem. Then, more costly patterns
can be activated on smaller parts of the program. This is similar to
what experienced programmers do. The difference with our approach
is that they do not have to either change tools, or reset the
parametrization of the debugger.

\section{Related Work}
\label{sec:analyse:related}

Kraut~\cite{bruegge83} implements a finite state machine to find
sequences of execution events that satisfy some patterns, called
\emph{path rules}. Several patterns are allowed and they can be
enabled or disabled during the execution, using a labeling policy.
Specified actions are triggered when a rule is satisfied but they are
limited to some debugger primitives, such as a message display or
incrementing a counter. The main interest of this tool is to abstract
the trace and to allow the easy development of monitors. The trace
analysis is necessarily synchronous and does not benefit from the
power of a complete programming language.

Reiss and Renieris~\cite{reiss01} have an approach similar to ours.
They also structure their dynamic analyses into three different
modules: 1) extraction of trace, 2) compaction and filtering and 3)
visualization. They provide a number of interesting compaction
functions which should be integrated in a further version of our
system. They, however, first dump the whole trace information in files
before any filtering is processed. With our tracer driver, filtering
is done on the fly, and Section~\ref{sec:analyse:performance} has
shown that this is much more efficient than first storing in
files. Their approach, however, is able to deal with partially ordered
execution threads; adapting our framework to languages with partially
ordered threads would require some technical work.

Coca~\cite{ducasse99} and Opium~\cite{ducasse99} provide a trace query
mechanism, respectively for C and Prolog. This mechanism is
synchronous and does not allow concurrent analyses. It can be easily
emulated with our tracer driver and an analyzer mediator written in
Prolog.

Hy$^+$~\cite{consens94} writes the trace into a real relational
database to query it with SQL. This is even slower than writing the
trace simply into a file. However, when on the fly performance is
not an issue, for example for post mortem analysis, this is a very
powerful and elegant solution which is straightforward to connect to
a tracer with our tracer driver.

Dalek~\cite{olsson90} is a powerful extension of \emph{gdb}. It allows
users to associate sequences of execution events to specific
synchronous handlers written in a dedicated imperative language. This
language includes primitives to retrieve additional trace data and to
synchronize the execution. The management of handlers is not
incremental. A key feature of Dalek, especially useful in an
imperative language, is the explicit ``queue of events'' that stores
the achieved execution events.  The user can explicitly remove events
from this queue and add higher-level events.  This approach requires
an expensive storage of a part of the trace but enables both
monitoring, debugging and profiling of programs.

In EBBA~\cite{bates95}, expected program behaviors are modeled as
relationships between execution events. Those models are then compared
to the actual behavior during execution. EBBA tries to recognize
relevant sequences of events and to check some constraints about such
sequences.  A kind of automaton is built to find instantiations of the
models.  The events are first generated by the tracer before being
filtered according to the automata. Our approach allows filtering
execution events directly inside the tracer, which is more efficient.
Nevertheless, EBBA recognizes sequences of events whereas we filter one
event at time. Our approach could be used upstream of the
sequence recognition. The incrementality of the event patterns could be
used to adapt the relevant events to the states of the automata.

UFO~\cite{auguston02} offers a more powerful language to specify
patterns and monitors than EBBA. The patterns can involve several
events, not necessarily consecutive.  In our framework, the monitors
have to be implemented in the analyzer with a general programming
language. A further extension should allow at least the implementation of
monitors in the trace driver to improve efficiency.  UFO, however, does
not allow the same flexibility as our tracer driver, and is heavier to
use for interactive debugging.

So far, our framework applies only to a single execution and does not
easily scale to compare numerous executions as is done in ``batch
mode'' by ~\cite{jones02}.  It seems, however, possible to extend our
framework so that two executions can be run in parallel with two
tracer drivers. This would allow the implementation of the debugging analyses
of~\cite{zeller02} and~\cite{sosic97} which compare two executions at
a given moment.

For some applications, it is important to be able to rewind
the execution. The necessary mechanisms are orthogonal to the ones
presented here, and can be merged with them. Interested readers are
referred to the Mercury mechanisms~\cite{maclarty06} or the survey
of~\cite{ronsse00c}.

The Ilog Christmas Tree~\cite{Outils:OPLstudio} is built by processing
an XML trace produced by the debugger. This tracer is generic: it can
be specialized to feed a specific tool. This specialization requires,
however, a good understanding of the solver behavior and cannot be
modified during the execution. Moreover, the amount of data available
at each event is very limited compared to the full state our approach allows. 
For instance, the set of constraints
and variables cannot be inspected.

A debugging library for SICStus Prolog has been
implemented~\cite{agren02,hanak04}.  Its main quality is the
explanations it provides about events that narrow domains. This
helpful information needs a difficult and costly instrumentation of
SICStus constraints: only a few ones have actually been instrumented.
No performance results are available.  Some tuning of the trace
display is possible but the tracer is based on a complete storage of
the trace and a postmortem investigation: this is impractical with
real-sized executions. The lazy generation of the trace our tracer
driver enables leads to the same kind of trace data in a more
efficient and practical way.

Some C(L)P debugging tools enable users to interact with the execution
states.  User of Oz Explorer~\cite{schulte97} can act on the current
state of the execution to drive the search-tree exploration. Users of
CLPGUI~\cite{fages02} can add new constraints on a partial
solution. They can recompute a former state in both systems.  Those
features are helpful. Our approach is complementary, it addresses the
communication from the traced execution to the debugging tools.

\section{Conclusion}

In this paper we have presented a tracer driver which, with limited
development efforts, provides a powerful front-end for complex debugging and
monitoring tools based on trace data.

We have defined an expressive language of event patterns where
relevant events are described by logical formul\ae{} involving most of
the data the tracer can access. Specific primitives enable the
retrieval of large pieces of data ``on demand'' and the adaptation of
the event patterns to the evolving needs of trace analyzers.

Experiments for CLP(FD) have shown that the overhead of the core
tracer mechanisms is small, therefore the core tracer can be
permanently activated;
the tracer driver overhead is acceptable;
there is no overhead in the filtering mechanisms when searching
simultaneously for several patterns;
the tracer driver overhead is predictable for given patterns;
the tracer driver approach that we propose is more
efficient than sending over a default trace, even to construct
sophisticated graphical views;
answering queries is orders of magnitude more efficient
than displaying traces;
there is no need to  restrict the trace information a priori;
last but not least, the performance of our tool is comparable to the
state-of-the-practice while being more powerful and more generic.

Traditionally, tracer designers decide on a static basis what the
observed events should be. As a result, compromises regarding the
amount of information to trace are made once for all or, at best,
before each execution. With our approach the trace contents can be
much richer because only what is needed is retrieved. Hence there is
less chance that important information is missing.

The tracer driver overhead is inversely proportional to the average
time between two traced events. Whereas the principles of the tracer
driver are independent of the traced programming language, it is best
suited for high-level languages, such as constraint logic programming,
where each traced execution event encompasses numerous low-level
execution steps. 

The current C(L)P environments do not provide all the useful dynamic
analysis tools. They can significantly benefit from our tracer driver
which enables dynamic analyses to be integrated  at a very low cost.

\begin{small}
\paragraph*{Acknowledgment}\quad
The authors thank Pierre Deransart and their \oadimpac{} partners for
fruitful discussions, as well as Guillaume Arnaud for the screenshots
of his Pavot tool and his careful beta-testing of Codeine. 
Ludovic Langevine is indebted to Mats Carlsson and the SICStus team at
Uppsala who helped to implement a preliminary tracer driver prototype
connected to the FDBG SICStus Prolog tracer~\cite{agren02}. It
contributed to the validation of the principles described in this
article.

\end{small}


\end{document}